\documentclass[aps,prl,twocolumn,superscriptaddress,showpacs]{revtex4-1}
\usepackage{amsmath}
\usepackage{color}
\usepackage{dcolumn}
\usepackage{amssymb}
\usepackage{epsfig}
\usepackage{multirow}
\usepackage{amsbsy}
\usepackage{array}
\usepackage{hyperref}
\usepackage{bm} 
\usepackage{extarrows}
\usepackage{graphicx}

\begin{document}

\title{Topological Superconductivity Intertwined with Broken Symmetries}
 
\author{Hui-Ke Jin}
\affiliation{Department of Physics, Zhejiang University, Hangzhou 310027, China}

\author{Dong-Hui Xu}
\email{donghuixu@hubu.edu.cn}
\affiliation{Department of Physics, Hubei University, Wuhan 430062, China}

\author{Yi Zhou}
\email{yizhou@iphy.ac.cn}
\affiliation {Beijing National Laboratory for Condensed Matter Physics $\&$ Institute of Physics, Chinese Academy of Sciences, Beijing 100190, China}
\affiliation{Department of Physics, Zhejiang University, Hangzhou 310027, China}
\affiliation{CAS Center for Excellence in Topological Quantum Computation, University of Chinese Academy of Sciences, Beijing 100190, China}
\affiliation{Collaborative Innovation Center of Advanced Microstructures, Nanjing University, Nanjing 210093, China}

\date{\today}

\begin{abstract}
Recently the superconductor and topological semimetal PbTaSe$_2$ was experimentally found to exhibit surface-only lattice rotational symmetry breaking below $T_c$.
We exploit the Ginzburg-Landau free energy and propose a microscopic two-channel model to study possible superconducting states on the surface of PbTaSe$_2$. 
We identify two types of topological superconducting states. One is time-reversal invariant and preserves the lattice hexagonal symmetry while the other breaks both symmetries. We find that such time-reversal symmetry breaking is unavoidable for a superconducting state in a two dimensional irreducible representation of crystal point group in a system where the spatial inversion symmetry is broken and the strong spin-orbit coupling is present. Our findings will guide the search for topological chiral superconductors.

\end{abstract}

\maketitle

Topological superconductors (TSCs) are novel quantum states of matter characterized by nontrivial topology of Cooper pairing states. Some of them are predicted to host Majorana zero modes (MZMs) in vortex cores~\cite{Read:00,Volovik:99,Ivanov:01} that obey non-Abelian braiding statistics and can be utilized as topological qubits in quantum computation~\cite{Kitaev:03,Nayak:08}. However, naturally occurring TSCs have been found rarely so far, people proposed an alternative approach to engineer TSC, namely, fabricating artificial heterostructures consisting of various systems in proximity to a conventional $s$-wave superconductor, such as three dimensional (3D) strong topological insulators (TIs)~\cite{Fu:08}, semiconductors with strong Rashba spin-orbit coupling (SOC) under an external Zeeman field~\cite{Sau:10,Lutchyn:10,Oreg:10,Alicea:10}, quantum anomalous Hall insulators~\cite{Qi:10}, and ferromagnetic atomic chains~\cite{Choy:11,Nadj-Perge:13}. Signatures of MZMs in these heterostructures have been reported experimentally~\cite{Mourik:12,Nadj-Perge:14,Deng:16,Sun:16}.

Meanwhile, the rapid development of topological materials shed new light on intrinsic TSCs. For instance, the chemical doping of TIs provides a promising route to TSC, by which a 3D TSC candidate Cu$_x$Bi$_2$Se$_3$ was discovered~\cite{Hor:10}. 
It was demonstrated by nuclear magnetic resonance (NMR) experiments that Cu$_x$Bi$_2$Se$_3$ is in an electronic nematic state below its superconducting transition temperature $T_c$~\cite{Matano:16}, i.e., the Knight shift exhibits two-fold symmetry when applied magnetic field is rotated in the $ab$-plane, even though the crystal lattice is hexagonal. 
This nematic superconducting state is consistent with the 2D $E_u$ representation of Cu$_x$Bi$_2$Se$_3$ crystal point group $D_{3d}$~\cite{Fu:10,Fu:14,Venderbos:16}. 
Subsequent measurements on Cu-, Nb-, and Sr-doped Bi$_2$Se$_3$ compounds display two-fold symmetry in the in-plane field-angle dependence of specific heat, magnetic torque, upper critical field,
magnetization, and scanning tunneling microscopy (STM) spectra in their superconducting states~\cite{Yonezawa:17,Asaha:17,Pan:16,Du:17,Andersen:18,Shen:17,Tao:18}. 

Another example is PbTaSe$_2$, which is a topological material on a non-centrosymmetric lattice with $P\bar{6}m2$ space group and will become superconducting below $T_c \sim 3.7$~K~\cite{Ali:14}.
Density functional theory (DFT) calculations and angle-resolved photoemission spectroscopy (ARPES) measurements reveal that its normal state has topological nodal rings in bulk in addition to an indirect gap opened by the SOC. The latter is characterized by a $\mathbb{Z}_2$ topological invariant, and associated surface states form two helical Fermi surfaces~\cite{Bian:16,Chang:16,Guan:16}.
The helical surface states have also been confirmed by scanning STM and quasiparticle interference (QPI) measurements, in which a full superconducting gap was observed~\cite{Guan:16} below $T_c$.
So far all the bulk measurements, including specific heat~\cite{Zhang:16}, thermal conductivity~\cite{Wang:16}, London penetration depth~\cite{Pang:16} and NMR~\cite{Maeda:18}, are consistent with the full gap scenario. And $\mu$SR experiments suggest that the time-reversal symmetry (TRS) is unbroken in the superconducting state~\cite{Wilson}.

Very recently, by magnetic-field-rotational heat capacity, resistivity and point-contact spectroscopy (PCS) measurements on single crystalline PbTaSe$_2$, Le et al.~\cite{Le:19} found that: (1) the resistivity measures an in-plane upper critical field $H_c^{R}\sim 1.0$~T, which is significantly larger than the corresponding value $H_{c}^{HC}\sim 0.2$~T measured by heat capacity and thermal conductivity; (2) the field-angle dependent soft PCS displays two-fold symmetry in the superconducting state despite its hexagonal lattice symmetry ($D_{3h}$ in bulk and $C_{3v}$ on surface), while this nematicity can not be observed by field-rotational specific heat; (3) such a reduced rotational symmetry occurs only in the superconducting state and will disappear with increasing temperature and be suppressed by a magnetic field exceeding $H_{c}^{R}$. Moreover, the nematicity exists in the field window $H_{c}^{HC}<H<H_{c}^{R}$, where the bulk is in the normal state while the surface is still a superconducting state.
This surface-only nematic superconductivity motivates us to study topological superconductors with broken symmetries.

In this Letter, we will start with studying possible topological superconductivity on the surface of PbTaSe$_2$ and then discuss how our results are applicable to more generic situations, e.g., the surface on Cu$_x$Bi$_2$Se$_3$ and the bulk in non-centrosymmetric superconductors. 

\begin{figure}[htpb]
	\includegraphics[scale=0.31]{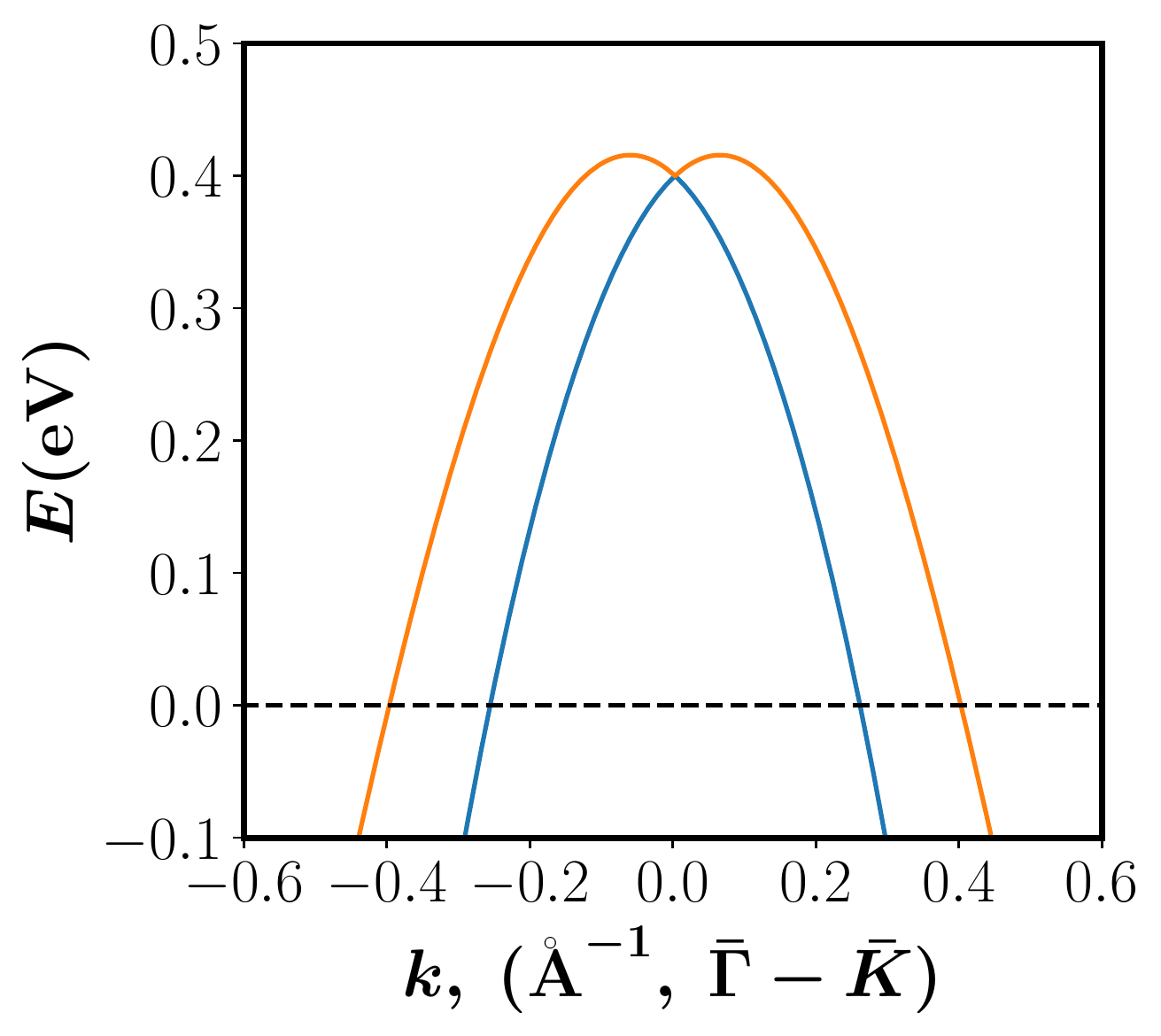}
	\includegraphics[scale=0.31]{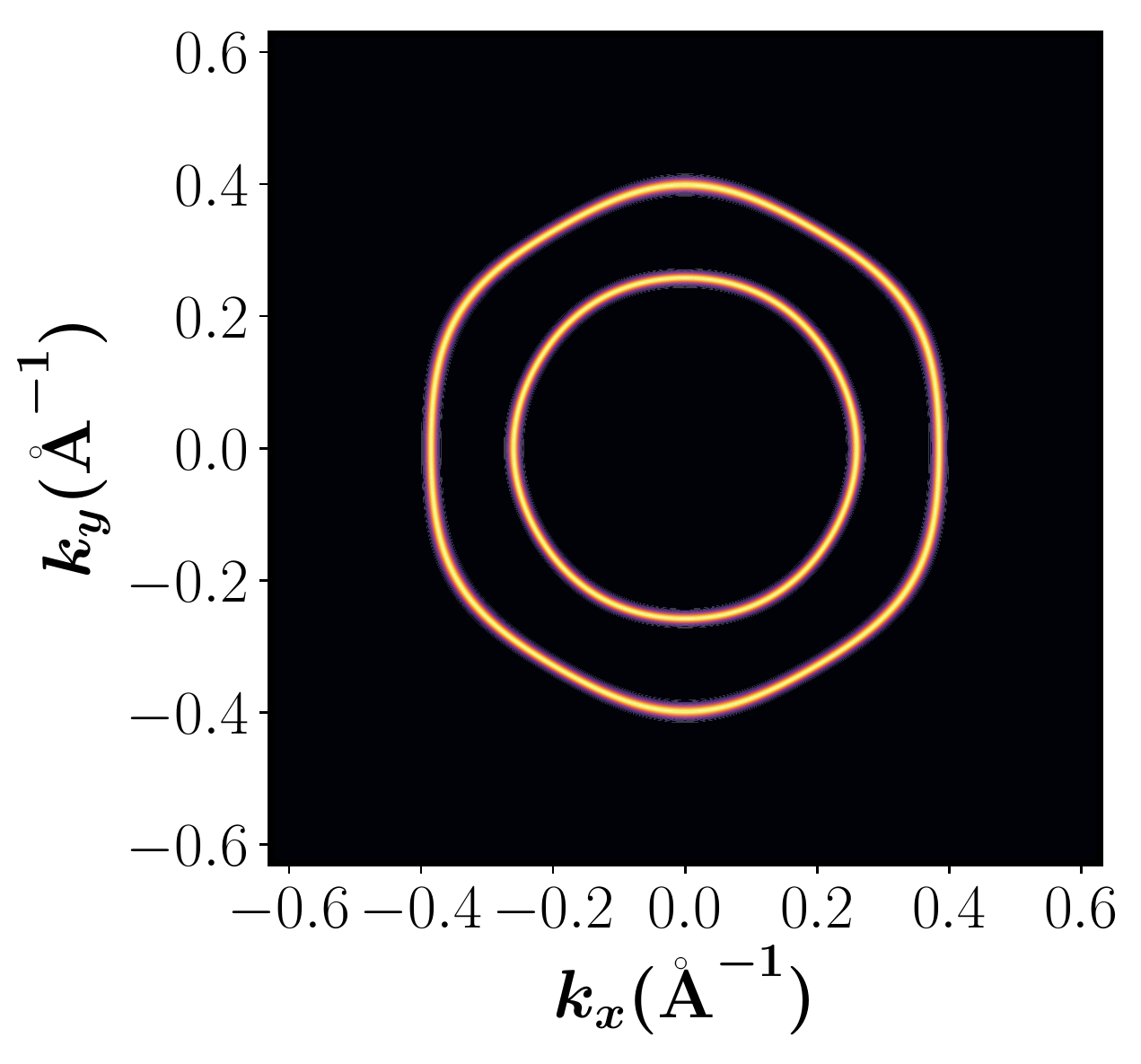}
	\caption{ Left: Energy dispersion of helical surface states along $\bar{\Gamma}$-$\bar{K}$ direction in the surface Brillouin zone. Right: The helical Fermi surfaces. }
	\label{fig:BSFS}
\end{figure}

\emph{Effective model for topological surface states.---} 
According to the DFT calculation and ARPES measurements on PbTaSe$_2$~\cite{Bian:16,Chang:16}, the topological surface states can be described by a $\bm{k}\cdot\bm{p}$ Hamiltonian near the $\bar{\Gamma}$ point, which to the third order in $\bm{k}$ takes the form $H_0=\sum_{\bm{k}}c^{\dagger}_{\bm{k}}H_0(\bm{k})c_{\bm{k}}$, where $c_{\bm{k}}=\left(c_{\bm{k}\uparrow},c_{\bm{k}\downarrow}\right)^{T}$ and
\begin{equation}\label{eq:H0}
H_0(\bm{k}) = \xi_{\bm{k}}\sigma_{0}-\alpha_{R}(k_{x}\sigma_{y}-k_{y}\sigma_{x})+\lambda_{w}(k_{+}^{3}+k_{-}^{3})\sigma_{z}. 
\end{equation}
$\xi_{\bm{k}}=\frac{k^{2}}{2m^{*}}-\mu$, $k^{2}=k_{x}^{2}+k_{y}^{2}$, $k_{\pm}=k_{x}\pm ik_{y}$, ${\sigma}_{x,y,z}$ are Pauli matrices for spins, and $\sigma_{0}$ is the identity matrix. $\alpha_{R}$ and $\lambda_{w}$ are the Rashba SOC and hexagonal wrapping strength respectively. The diagonalization of $H_{0}(\bm{k})$ gives rise to two helical eigenstates with the energy spectrum, $\epsilon_{\bm{k}\pm}= \xi_{\bm{k}} \pm{}\sqrt{\alpha_{R}^{2}k^{2}+\lambda_{w}^{2}(k_{+}^{3}+k_{-}^{3})^{2}}$.
In this study, we choose the parameters as follows: $m^{*}=-0.125~\mbox{eV}^{-1}\mbox{\r{A}}^{-2}$, $\alpha_{R}=0.5~\mbox{eV}\mbox{\r{A}}$, $\lambda_{w}=1~\mbox{eV}\mbox{\r{A}}^{3}$ and $\mu=-0.4~\mbox{eV}$, which reproduces the energy dispersion along $\bar{\Gamma}-\bar{K}$ direction and two Fermi surfaces as illustrated in Fig.~\ref{fig:BSFS}.

\emph{Superconducting pairing symmetry.---}
Now we consider superconducting pairing for these topological surface states. The inversion symmetry is broken on the surface, such that a Cooper paired state will be an admixture of spin-singlet and -triplet in the presence of SOC. To begin with and for later purpose, we list possible pairing functions on a 2D $C_{3v}$ lattice in Table \ref{tab:basis}, where even-parity and spin-singlet gap functions $\psi(\bm{k})$ and odd-parity and spin-triplet gap functions $\bm{d}(\bm{k})$ are classified in accordance with $C_{3v}$ group irreducible representation $\Gamma$. Note that all the $k_z$ relevant terms vanish on the surface.

\begin{table}[hbtp]
	\caption{Irreducible representations of $C_{3v}$ point group and corresponding even and odd-parity basis functions.}\label{tab:basis}
	\renewcommand\arraystretch{1.75}
	\setlength{\tabcolsep}{1.5mm}{
		\begin{tabular}{ccc}
			\hline
			\hline
			$\Gamma$ & $\psi(\bm{k})$ & $\bm{d}(\bm{k})$\\
			\hline
			$A_{1}$  & $\psi^{A_{1}}$=1, $\psi^{A_{1}}_{\bm{k}}=k_{x}^{2}+k_{y}^{2}$ & $\bm{d}^{A_{1}}_{\bm{k}}=k_{x}\bm{\mathrm{x}}+k_{y}\bm{\mathrm{y}}$\\
			$A_{2}$  &                          & $\bm{d}^{A_{2}}_{\bm{k}}=k_{x}\bm{\mathrm{y}}-k_{y}\bm{\mathrm{x}}$ \\
			$E$      & $\psi^{E}_{1,\bm{k}}=k_{x}^{2}-k_{y}^{2}$ & $(\bm{d}^{E,\mathrm{z}}_{1,\bm{k}},\bm{d}^{E,\mathrm{z}}_{2,\bm{k}})=(k_{x},k_{y})\bm{\mathrm{z}}$\\
			& $\psi^{E}_{2,\bm{k}}=2k_{x}k_{y}$ &  $\bm{d}^{E,\mathrm{xy}}_{1,\bm{k}}=k_{x}\bm{\mathrm{y}}+k_{y}\bm{\mathrm{x}}$\\
			& & $\bm{d}^{E,\mathrm{xy}}_{2,\bm{k}}=k_{x}\bm{\mathrm{x}}-k_{y}\bm{\mathrm{y}}$\\
			\hline
			\hline
	\end{tabular}}
\end{table}

In particular, we are interested in the $E$ representation, because it is the only one representation that is able to give rise to a nematic state. 
Without loss of generality, a two-component superconducting pairing function in the $E$ representation reads,
\begin{equation}\label{eq:pairing}
\Delta_{\bm{k}} = \phi_{1} \Delta_{1\bm{k}} + \phi_{2} \Delta_{2\bm{k}},
\end{equation}
where $\phi_{m=1,2}$ are two complex numbers labeling order parameters associated with the two components, and the two basis functions $\Delta_{n\bm{k}}$ take the generic form
$\Delta_{m\bm{k}}=i\left(\psi_{m\bm{k}}\sigma_{0}+\bm{d}_{m\bm{k}}\cdot\bm{\sigma}\right)\sigma_{y}$. Here $\bm{\sigma}$ consists of three components of Pauli matrices, $\psi_{m\bm{k}}$ and $\bm{d}_{m\bm{k}}$ are given in Table \ref{tab:basis}.
By the Wigner's theorem \cite{Wigner}, $\psi_{m}(\bm{k})$ and $\bm{d}_{m\bm{k}}$ can be chosen as real functions, since they arise from two-electron pairing. 
In this basis, the Ginzburg-Landau free energy on a $C_{3v}$ lattice can be formulated up to the quadratic order:
\begin{eqnarray}
\mathcal{F}&=&\alpha(T-T_{c})(|\phi_{1}|^{2}+|\phi_{2}|^{2}) + \beta_{1}\left(|\phi_{1}|^{2}+|\phi_{2}|^{2}\right)^{2} \nonumber\\
& & + \beta_{2}\left(\phi_{1}\phi_{2}^{*}-\phi_{1}^{*}\phi_{2}\right)^{2}. \label{eq:GL}
\end{eqnarray}
The coefficient $\alpha>0$ gives rise to superconducting order at $T<T_{c}$. And it can be justified that $\beta_{1}>0$ so that the system is stable. Note that the free energy in Eq.~\eqref{eq:GL} shares the same form with that on a $D_6$ lattice~\cite{Sigrist:91}, and is applicable equally to single-band and multi-band. It turns out that the sign of $\beta_{2}$ is crucial to the nematicity: (1) if $\beta_{2}>0$, $\phi_{1}$ and $\phi_{2}$ develop a relative $\pi/2$ phase, say, $\left(\phi_{1},\phi_{2}\right)$=$\phi\left(1,\pm i\right)$, resulting in a chiral superconductor; (2) while if $\beta_{2}<0$, $\left(\phi_{1},\phi_{2}\right)$=$\phi\left(1,\pm 1\right)$ and nematic states are energetically favored \cite{Venderbos:16,Huang:18}. The coefficient $\beta_2$ depends on the structure of $\psi_{n\bm{k}}$ and $\bm{d}_{n\bm{k}}$ as well as the normal state Hamiltonian.

\emph{The effect of broken inversion symmetry.---} To see the effect of broken inversion symmetry and for the purpose of comparison, we consider a normal state with inversion symmetry and TRS temporarily, for which the double degenerate energy band will not split even though the effect of SOC is accounted. Replace spins by pseudospins, the superconducting pairing now will reduce to either pseudospin-singlet or -triplet, and the coefficient $\beta_2$ will be always positive for singlet states and read  $\beta_2\simeq C \left\langle (\bm{d}_{1}\cdot\bm{d}_{2})^{2}-|\bm{d}_{1}\times{}\bm{d}_{2}|^{2} \right\rangle_{\text{FS}}$ for triplet states, 
where $C$ is positive and $\left\langle\cdots\right\rangle_{\text{FS}}$ is an average over the Fermi surface (see the Supplementary Materials).

The situation will change dramatically in the absence of inversion symmetry. 
It will be more appropriate to express the pairing function $\Delta_{\bm{k}}$ in the pseudospin basis in the presence of SOC splitting. Namely, the pairing terms can be rewritten as $\Delta_{ss^{\prime}}(\bm{k})c_{\bm{k}s}^{\dagger}c_{-\bm{k}s^{\prime}}^{\dagger}$, where $s,s^{\prime}=\pm$ refer to upper or lower helical band. When the SOC splitting $\Delta_{\text{SOC}}$ is much larger than the superconductor gap $\Delta_{\text{SC}}$, the intra-band superconducting pairing will be energetically favored rather than the inter-band pairing, which can be precisely summarized by the following Lemma.

{\bf Lemma}: For a superconducting system with strong SOC splitting, say, $\Delta_{\text{SOC}}\gg \Delta_{\text{SC}}$, the Ginzburg-Landau free energy is of the form,
\begin{equation}
\mathcal{F} = \sum_{\bm{k}}\sum_{s=\pm}F_{s}[\Delta_{ss}(\bm{k})] + O\left(\frac{\Delta_{\text{SC}}^2}{\Delta_{\text{SOC}}}\right).
\end{equation}
The proof and the exact forms of function $F_{s}(x)$ can be found in the Supplementary Materials.
With the help of this Lemma, one can prove a symmetry-breaking theorem as follows.

{\bf Theorem}: For a lattice with broken inversion symmetry, consider the superconducting pairing functions in a 2D irreducible representation given by Eq.~\eqref{eq:pairing}, the two order parameters $\phi_1$ and $\phi_2$ will develop a relative $\pi/2$ phase and break TRS when $\Delta_{\text{SOC}}\gg \Delta_{\text{SC}}$.

{\bf Proof}: Begin with a lattice of $C_{3v}$ symmetry, the free energy now can be written in terms of $\Delta_{ss}(\bm{k})$ as follows, $\mathcal{F} = \alpha (T-T_{c}) (|\phi_{1}|^2+|\phi_{2}|^2) + \sum_{\bm{k},s=\pm}|g_{s}(\bm{k})|^{4}|\Delta_{ss}(\bm{k})|^4 + O\left(\frac{\Delta_{\text{SC}}^2}{\Delta_{\text{SOC}}}\right)$, where $g_{s}(\bm{k})$ is given by the normal state $\bm{k}\cdot\bm{p}$ Hamiltonian, $\Delta_{ss}({\bm{k}}) = \phi_{1} \Delta_{1ss}(\bm{k}) + \phi_{2} \Delta_{2ss}(\bm{k})$, and $\Delta_{nss(n=1,2)}$ are two basis functions in the $E$ representation. Thus the coefficient $\beta_2$ in Eq.~\eqref{eq:GL} reads $\beta_2=\sum_{\bm{k},s=\pm}|g_{s}(\bm{k})|^4\Delta_{1ss}(\bm{k})^2\Delta_{2ss}(\bm{k})^{*2} $. The Wigner's theorem \cite{Wigner} tells us that $\Delta_{1ss}$ and $\Delta_{2ss}$ can be always chosen as real functions. So that $\beta_{2}>0$ and $(\phi_1,\phi_2)=\phi(1,\pm i)$ will minimize $\mathcal{F}$ subject to the constraint $|\phi_1|^2+|\phi_2|^2=2|\phi|^2$. It is easy to verify that the above proof is applicable to other crystal point groups with 2D irreducible representations too, including $D_{2d}$, $D_4$, $C_{4v}$, $D_3$, $D_6$, $C_{6v}$, $D_{3h}$, $T$, $O$ and $T_{d}$. QED. It is worth noting that a similar conclusion for spinless fermions was obtained in Ref. \cite{Cheng:10}.

\emph{Projected pairing functions.---}
It will be more physically transparent to illustrate how a Cooper paired state in the spin basis is projected to the pseudospin basis~\cite{Fu:08,Alicea:10}, and how a projected state transfers under time reversal (TR) in the pseudospin basis. In Table~\ref{tab:Pgap}, we summarize projected pairing functions in the pseudospin basis in accordance with those in the spin basis given in Table~\ref{tab:basis}.
In regard to TR operation $\mathcal{T}$, we have $\mathcal{T}c_{\bm{k}\uparrow}^{\dagger}\mathcal{T}^{-1}=c_{-\bm{k}\downarrow}^{\dagger}$ and $\mathcal{T}c_{\bm{k}\downarrow}^{\dagger}\mathcal{T}^{-1}=-c_{-\bm{k}\uparrow}^{\dagger}$ in the spin basis, so that the operators in the pseudospin basis will transfer as:
$\mathcal{T}c_{\bm{k}\pm}^{\dagger}\mathcal{T}^{-1}  =  \mp\mbox{sgn}(\alpha_{R})e^{\mp{}i\theta_{\bm{k}}}c_{-\bm{k}\pm}^{\dagger}$,
$\mathcal{T}c_{\bm{k}\pm}^{\dagger}c_{-\bm{k}\pm}^{\dagger}\mathcal{T}^{-1}  =  e^{\mp{}2i\theta_{\bm{k}}}c_{\bm{k}\pm}^{\dagger}c_{-\bm{k}\pm}^{\dagger}$,
$\mathcal{T}c_{\bm{k}\pm}^{\dagger}c_{-\bm{k}\mp}^{\dagger}\mathcal{T}^{-1} =  - c_{\bm{k}\pm}^{\dagger}c_{-\bm{k}\mp}^{\dagger}$,
where $\theta_{\bm{k}}$ is the azimuthal phase defined by $(k_x,k_y)=k(\cos\theta_{\bm{k}},\sin\theta_{\bm{k}})$.
Therefore a TR invariant pairing function can be written explicitly as $\Delta_{\pm\pm}=ie^{\mp i\theta_{\bm{k}}}f_{\pm\pm}(\bm{k})$ and $\Delta_{\pm \mp}=if_{\pm \mp}(\bm{k})$, where $f_{ss^{\prime}}^{*}(-\bm{k})=f_{ss^{\prime}}(\bm{k})$. All the states in Table ~\ref{tab:Pgap} are TR invariant, and the broken inversion symmetry and strong SOC will give rise to TRS breaking in the $E$ representation and result in TSCs.

\begin{table*}[htbp]
\caption{Projected pairing functions in the pseudospin basis ($\Delta_{\pm\pm}$ and $\Delta_{\pm\mp}$) in accordance to those in the spin basis ($\psi(\bm{k})$ and $\bm{d}(\bm{k})$). The factors $u_{\bm{k}}$ and $v_{\bm{k}}$ are determined by the normal state Hamiltonian given in Eq.~\eqref{eq:H0}, and read $u_{\bm{k}}$=$|\alpha_{R}k|/\sqrt{\alpha_{R}^{2}k^{2}+V_{\bm{k}}^{2}}$ and $v_{\bm{k}}$=$\mbox{sgn}(\alpha_{R})V_{\bm{k}}/\sqrt{\alpha_{R}^{2}k^{2}+V_{\bm{k}}^{2}}$, where $V_{\bm{k}}=\lambda_{w}(k_{+}^{3}+k_{-}^{3})$.
}\label{tab:Pgap}
\renewcommand\arraystretch{1.2}
\setlength{\tabcolsep}{2mm}{
\begin{tabular}{c|c|c|c|c|c}
	\hline
	\hline
	IR & $\psi(\bm{k})$ or $\bm{d}(\bm{k})$ &  $\Delta_{++}$   & $\Delta_{+-}$   &  $\Delta_{-+}$ &  $\Delta_{--}$\\
	\hline
	$A_{1}$ & $1$ &  $i{}e^{-i\theta_{\bm{k}}}$ & $0$ & $0$ & $-i{}e^{i\theta_{\bm{k}}}$ \\
	\hline
	$A_{1}$ & $\cos\theta_{\bm{k}}\bm{\mathrm{x}}+\sin\theta_{\bm{k}}\bm{\mathrm{y}}$ &  $0$ & $-i$ & $i$ & $0$ \\
	\hline
	$A_{2}$ & $\cos\theta_{\bm{k}}\bm{\mathrm{y}}-\sin\theta_{\bm{k}}\bm{\mathrm{x}}$ &  $iu_{\bm{k}}e^{-i\theta_{\bm{k}}}$ & $v_{\bm{k}}$ & $v_{\bm{k}}$ & $iu_{\bm{k}}e^{i\theta_{\bm{k}}}$ \\
	\hline
	& $\sin(2\theta_{\bm{k}})$ &  $i\sin(2\theta_{\bm{k}})e^{-i\theta_{\bm{k}}}$ & $0$ & $0$ & $-i\sin(2\theta_{\bm{k}})e^{i\theta_{\bm{k}}}$ \\
	\cline{2-6}
	\raisebox{2.3ex}[0pt]{$E$} & $\cos(2\theta_{\bm{k}})$ &  $i\cos(2\theta_{\bm{k}})e^{-i\theta_{\bm{k}}}$ & $0$ & $0$ & $-i\cos(2\theta_{\bm{k}})e^{i\theta_{\bm{k}}}$ \\
	\hline
	& $\cos\theta_{\bm{k}}\bm{\mathrm{z}}$   & $-iv_{\bm{k}}\cos\theta_{\bm{k}}e^{-i\theta_{\bm{k}}}$ & $u_{\bm{k}}\cos\theta_{\bm{k}}$ & $u_{\bm{k}}\cos\theta_{\bm{k}}$ & $-iv_{\bm{k}}\cos\theta_{\bm{k}}e^{i\theta_{\bm{k}}}$ \\
	\cline{2-6}
	\raisebox{2.3ex}[0pt]{$E$}  & $\sin\theta_{\bm{k}}\bm{\mathrm{z}}$  & $-iv_{\bm{k}}\sin\theta_{\bm{k}}e^{-i\theta_{\bm{k}}}$ & $u_{\bm{k}}\sin\theta_{\bm{k}}$ & $u_{\bm{k}}\sin\theta_{\bm{k}}$ & $-iv_{\bm{k}}\sin\theta_{\bm{k}}e^{i\theta_{\bm{k}}}$ 
	\\
	\hline
	& $\cos\theta_{\bm{k}}\bm{\mathrm{x}}-\sin\theta_{\bm{k}}\bm{\mathrm{y}}$ & $-iu_{\bm{k}}\sin(2\theta_{\bm{k}})e^{-i\theta_{\bm{k}}}$ & $-i\cos(2\theta_{\bm{k}})-v_{\bm{k}}\sin(2\theta_{\bm{k}})$ & $i\cos(2\theta_{\bm{k}})-v_{\bm{k}}\sin(2\theta_{\bm{k}})$ & $-iu_{\bm{k}}\sin(2\theta_{\bm{k}})e^{i\theta_{\bm{k}}}	$ \\
	\cline{2-6}
	\raisebox{2.3ex}[0pt]{$E$}  & $\cos\theta_{\bm{k}}\bm{\mathrm{y}}+\sin\theta_{\bm{k}}\bm{\mathrm{x}}$ &  $-iu_{\bm{k}}\cos(2\theta_{\bm{k}})e^{-i\theta_{\bm{k}}}$ & $-i\sin(2\theta_{\bm{k}})+v_{\bm{k}}\cos(2\theta_{\bm{k}})$ & $i\sin(2\theta_{\bm{k}})+v_{\bm{k}}\cos(2\theta_{\bm{k}})$ & $-iu_{\bm{k}}\cos(2\theta_{\bm{k}})e^{i\theta_{\bm{k}}}	$ \\
	\hline
	\hline
\end{tabular}
}
\end{table*}

\emph{A two-channel model.---}
To diagnose the microscopic origination of the surface nematic superconductivity, we exploit the following linearized gap equation to calculate the transition temperature $T_{c}$ for various pairing states \cite{Sigrist:91},
\begin{equation}\label{eq:lgapeq}
\begin{split}
\Delta_{s_{1}s_{2}}(\bm{k})&=-T_{c}\sum_{i\omega_{l}s_{3}s_{4}\bm{q}}V_{s_{1}s_{2}}^{s_{3}s_{4}}(\bm{k},\bm{q})\\
&\times\left[G(i\omega_{l},\bm{q})\Delta(\bm{q})G(-i\omega_{l},-\bm{q})\right]_{s_{3}s_{4}},
\end{split}
\end{equation}
where $s_{1-4}$ denote spins and $G(i\omega_{l},\bm{q})$ is the normal-state Matsubara Green's function. $V_{s_{1}s_{2}}^{s_{3}s_{4}}(\bm{k},\bm{q})$ describes a two-channel interaction and can be written in terms of the basis functions given in Table~\ref{tab:basis},
\begin{equation}\label{eq:Veff}
\begin{split}
V_{s_{1}s_{2}}^{s_{3}s_{4}}(\bm{k},\bm{q}) = &-{g_{\text{on}}}\tilde{\psi}^{A_{1}}\tilde{\psi}^{A_{1}*}(i\sigma_{y})_{s_{1}s_{2}}(i\sigma_{y})^{\dagger}_{s_{3}s_{4}}\\
-&{g_{\text{nn}}}\sum_{\Gamma,m}\tilde{\psi}^{\Gamma}_{m,\bm{k}}\tilde{\psi}^{\Gamma}_{m,\bm{q}}(i\sigma_{y})_{s_{1}s_{2}}(i\sigma_{y})^{\dagger}_{s_{3}s_{4}}\\
-&{g_{\text{nn}}}\sum_{\Gamma,m}(i\tilde{\bm{d}}^{\Gamma}_{m,\bm{k}}\cdot\bm{\sigma}\sigma_{y})_{s_{1}s_{2}}(i\tilde{\bm{d}}^{\Gamma}_{m,\bm{q}}\cdot\bm{\sigma}\sigma_{y})^{\dagger}_{s_{3}s_{4}},
\end{split}
\end{equation}
where $g_{\text{on}}$ and $g_{\text{nn}}$ are the coupling strengths in on-site and nearest neighboring inter-site channels respectively, and $g_{\text{on}},g_{\text{nn}}>0\,(<0)$ stands for attractive (repulsive) interactions. 
Here $\tilde{\psi}^{\Gamma}_{m,\bm{k}}$ and $\tilde{\bm{d}}^{\Gamma}_{m,\bm{k}}$ are normalized basis functions (at each Fermi surface) corresponding to $\psi^{\Gamma}_{m,\bm{k}}$ and $\bm{d}^{\Gamma}_{m,\bm{k}}$  in Table~\ref{tab:basis}. 

\begin{figure}[htbp]
	\includegraphics[width=8.4cm]{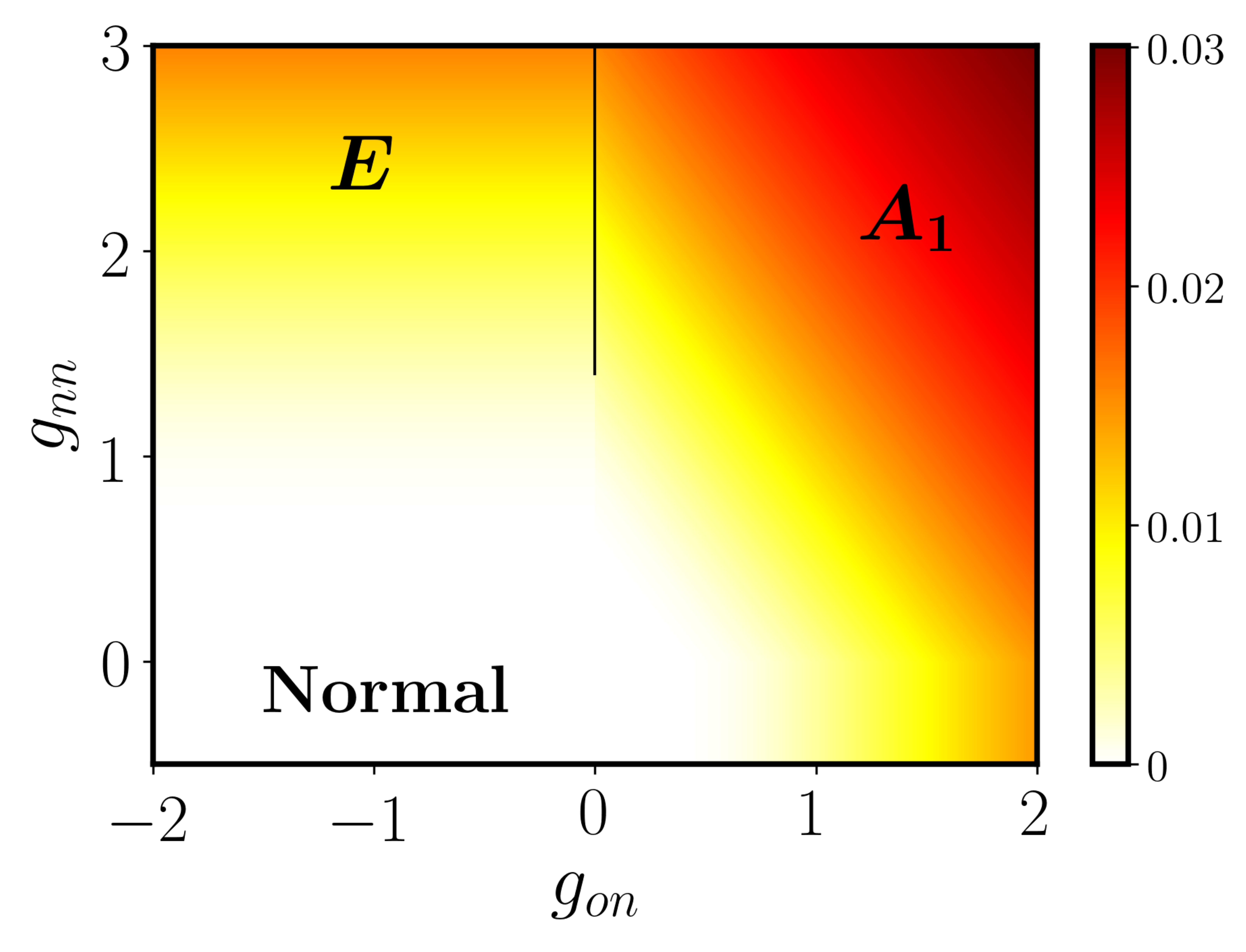}
	\caption{ Phase diagram for the two-channel $g_{\text{on}}$-$g_{\text{nn}}$ model. }\label{fig:phasediagram}	
\end{figure}

By setting a cutoff $\hbar\omega_{c}=0.05$~eV, one is able to solve Eq.~\eqref{eq:lgapeq} numerically to obtain $T_{c}$ for a given gap function $\Delta(\bm{k})$. 
With the help of Table~\ref{tab:basis}, the possible gap functions can be found as follows: 
$\Delta(\bm{k})|_{\Gamma=A_{1}}= -i(s_{A_{1}}\psi^{A_{1}} + s_{A_{1}}^{\prime}\psi^{A_{1}}_{\bm{k}} + t_{A_{1}}\bm{d}^{A_{1}}_{\bm{k}}\cdot\bm{\sigma})\sigma_{y}$,
$\Delta(\bm{k})|_{\Gamma=A_{2}}= -i(\bm{d}^{A_{2}}_{\bm{k}}\cdot\bm{\sigma})\sigma_{y}$,
and $\Delta(\bm{k})|_{\Gamma=E}=  \sum_{m=1,2}-i(s_{E,m}\psi^{E}_{m,\bm{k}} +t^{\mathrm{z}}_{E,m}\bm{d}^{E,\mathrm{z}}_{m,\bm{k}}\cdot\bm{\sigma} +t^{\mathrm{xy}}_{E,m}\bm{d}^{E,\mathrm{xy}}_{m,\bm{k}}\cdot\bm{\sigma})\sigma_{y}$,
where $s_{\Gamma}$'s and $t_{\Gamma}$'s are the coefficients of spin-singlet and spin-triplet components respectively.
For $A_1$ and $E$ states, the $T_{c}$ can be further optimized through varying the coefficients $s_{\Gamma}$'s and $t_{\Gamma}$'s.

The phase diagram for the two-channel $g_{\text{on}}$-$g_{\text{nn}}$ model is determined by the highest $T_{c}$ states and is plotted in Fig.~\ref{fig:phasediagram}. There are three phases: $E$, $A_1$ and the normal phase. 
(1) When $g_{\text{on}}>0$, the $s$-wave pairing $\psi^{A_{1}}$ dominates and the highest $T_{c}$ always arises in an $A_{1}$ state, which consists of negligible $p$-wave pairing $\bm{d}^{A_{1}}_{\bm{k}}$. Since the latter only gives rise to inter-band pairing (see Table~\ref{tab:Pgap}) and will be suppressed by the strong SOC according to the Lemma.
(2) When $g_{\text{on}}<0$ and $g_{\text{nn}}>0$, the pairing states in the $E$ representation will be favored.
These TRS invariant and breaking $E$ states are degenerate at $T_c$ due to the linearity, while the degeneracy will be lifted at lower temperatures and the ground states will break TRS due to our symmetry-breaking theorem. Moreover, these states are all admixture of spin-singlet and spin-triplet pairing.
(3) For $g_{\text{on}}<0$ and $g_{\text{nn}}<0$, we always have $T_{c}=0$, thereby the ground state is a normal state. 
It is worth nothing that both $A_1$ and $E$ phase are topologically nontrivial, which can be characterized by the winding number in each band, $N_{s} = \frac{1}{2\pi}\oint_{\text{FS}} d\bm{k}\cdot \nabla_{\bm{k}}\arg \Delta_{ss}(\bm{k})$. For an $A_1$ state, $N_{\pm}=\mp 1$; while for an $E$ state, either $N_{+}=1,N_{-}=3$ or $N_{+}=-3,N_{-}=-1$.

\emph{Quasiparticle gap.---}
Now we study the quasiparticle excitation gap in the $E$ phase. For comparison, we would like to consider more generic pairing states with $\Delta(\bm{k})$=$-i\sum_{m=1}^{2}\phi_{m}[\epsilon_{1}\psi^{E}_{m,\bm{k}}+(\epsilon_{2}\bm{d}^{E,\mathrm{z}}_{m,\bm{k}}+\epsilon_{3}\bm{d}^{E,\mathrm{xy}}_{m,\bm{k}})\cdot\bm{\sigma}]\sigma_{y}$ in addition to the pairing states derived from the two-channel model,
where $\epsilon_{1,2,3}$ are real parameters, and $(\phi_{1},\phi_{2})=\phi(1,1)$ or $\phi(1,i)$ depending on whether the system breaks TRS. 
If the system is TR invariant, i.e., $(\phi_{1},\phi_{2})=\phi(1,1)$, the quasiparticle gap is always nodal; otherwise, for $(\phi_{1},\phi_{2})=\phi(1,i)$ the gap is generally nodeless.
{\em It is found that the excitation gap will be of two-fold rotational symmetry as long as the pairing is an admixture of spin-singlet and spin-triplet.}

\begin{figure}
	\includegraphics[width=8.4cm]{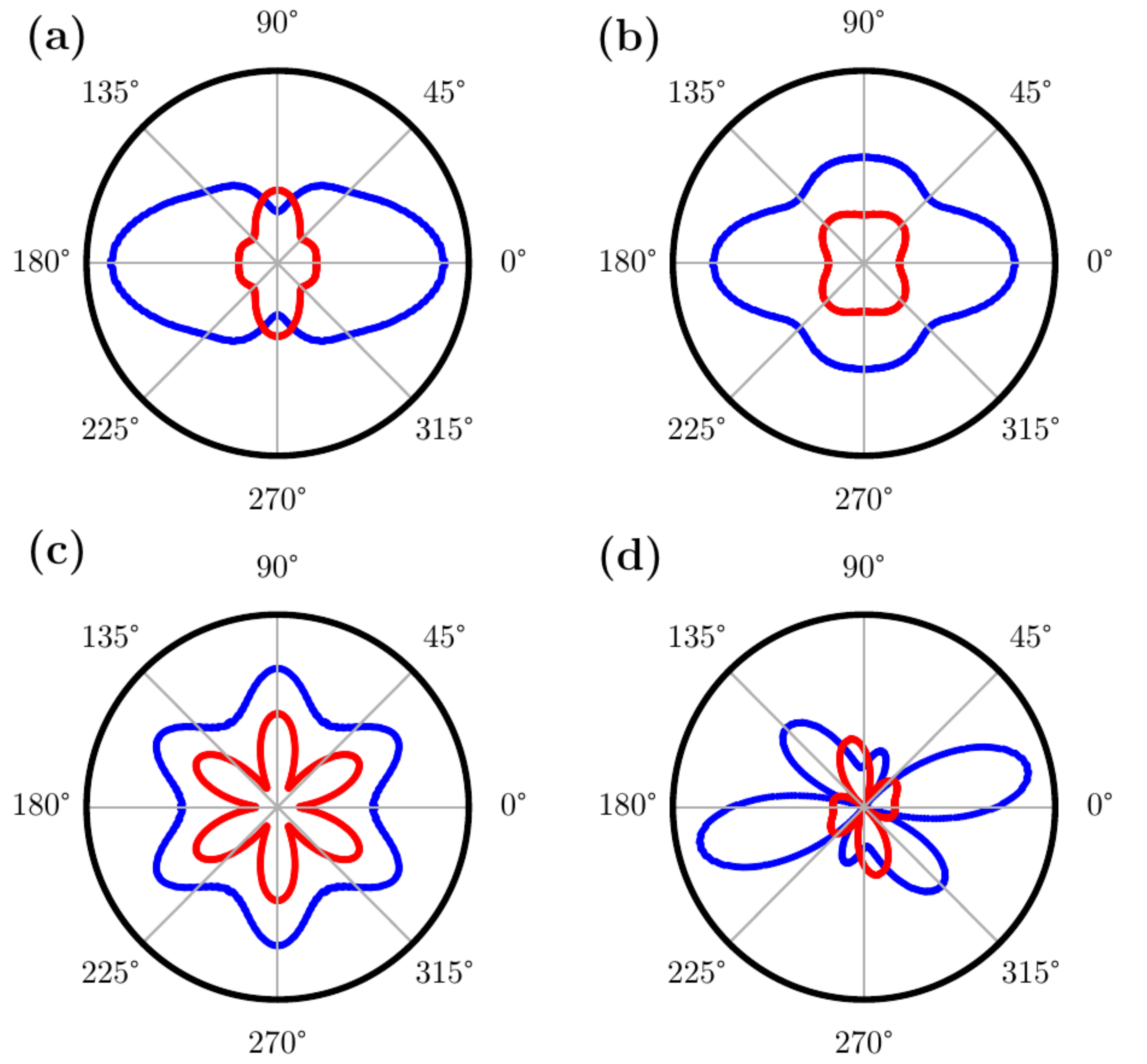}
	\caption{Quasiparticle gap of pairing states in the $E$ representation in the polar coordinate. Blue lines denote the gap at the outer Fermi surface (upper band, see FIG.~\ref{fig:BSFS}) and red lines denote the gap at the inner Fermi surface (lower band). (a)-(c) TRS breaking states with $(\phi_{1},\phi_{2})$=$\phi(1,i)$, and (d) a TR invariant state with $(\phi_{1},\phi_{2})$=$\phi(1,1)$. The parameters $\epsilon_{1,2,3}$ are chosen as:  
		(a) $\epsilon_{1}\approx\epsilon_{3}\approx\epsilon_{2}/3$ (a typical set of values derived from the two-channel model),
		(b) $\epsilon_{1}=\epsilon_{2}$, $\epsilon_{3}=0$,	
		(c) $\epsilon_{1}=0$, $\epsilon_{2}=2\epsilon_{3}$ (a pure spin-triplet state),
		(d) $\epsilon_{1}\approx\epsilon_{3}\approx\epsilon_{2}/3$.
	}\label{fig:qstruc}
\end{figure}

Several representative gap structures are illustrated in FIG.~\ref{fig:qstruc}.  A typical pairing state in the $E$ phase of the two-channel model is given by $(\phi_{1},\phi_{2})=\phi(1,i)$ and $\epsilon_{1}\approx\epsilon_{3}\approx\epsilon_{2}/3$, whose gap structure is plotted in FIG.~\ref{fig:qstruc}(a). Other two TRS breaking states are shown in FIG.~\ref{fig:qstruc} (b) and (c) as well. One consists of both spin-singlet and spin-triplet and is a nematic state (b), and the other is a pure spin-triplet pairing state and the excitation gap exhibits hexagonal symmetry (c).  
A TR invariant state with gap nodes is demonstrated in FIG.~\ref{fig:qstruc}(d).

\emph{Summary and discussions.---}
In summary, we theoretically studied the superconductivity of the topological surface states in PbTaSe$_2$. With the help a two-channel $g_{\text{on}}$-$g_{\text{nn}}$ model, we find that two types of superconducting pairing states are favored, which are in $A_{1}$ and $E$ representation respectively, and both of them are topologically non-trivial. The $A_1$ states are TRS invariant and do not break lattice hexagonal symmetry as the same as the TSC emerging in the 3D TI/$s$-wave superconductor heterosturcture~\cite{Fu:08}. In contrast, the $E$ states break TRS and lattice hexagonal symmetry resulting in TSCs with nematicity. We find that the TRS breaking in the $E$ states is unavoidable as long as the system breaks spatial inversion symmetry and the strong SOC is present. This symmetry breaking theorem is applicable to any crystal symmetries with 2D irreducible representation. One prediction is that the superconducting state on the surface of Cu$_{x}$Bi$_2$Si$_3$ must break TRS, which can be detected by the Kerr rotation. The other deduction is that if nematicity is observed in a non-centrosymmetric superconductor the TRS must break simultaneously.

\emph{Acknowledgement.}
We would like to thank Xin Lu, Shiyan Li, Xiangang Wan, Xiaofeng Xu, Hong Yao, and Fu-Chun Zhang for helpful discussions.
This work is supported in part by National Key Research and Development Program of China (No.2016YFA0300202),
National Natural Science Foundation of China (No. 11774306, No. 11704106), and the Strategic Priority Research Program of Chinese Academy of Sciences (No. XDB28000000). D.H.X. also acknowledges the financial support of the Chutian Scholars Program in Hubei Province.

\appendix

\begin{widetext}

\section{Gingzburg-Landau theory: the coefficient $\beta_2$ for a centrosymmetric superconductor}

The Gorkov Green's function in a superconducting system reads
\begin{equation}
\mathcal{G}^{-1}(i\omega_{l},\bm{k}) \equiv \mathcal{G}_{0}^{-1}(i\omega_{l},\bm{k}) + \Sigma(\bm{k}),
\end{equation}
where 
\begin{equation}
\mathcal{G}_{0}^{-1}(i\omega_{l},\bm{k})=\left(\begin{array}{cc}
G^{-1}(i\omega_{l},\bm{k}) & 0\\
0 & -G^{-1}(-i\omega_{l},-\bm{k}) \\
\end{array}\right),\ 
\Sigma(\bm{k})=\left(\begin{array}{cc}
0 & \Delta_{\bm{k}}\\
\Delta_{\bm{k}}^{\dagger} & 0 \\
\end{array}\right).
\end{equation}
Here $G^{-1}(i\omega_{l},\bm{k})=(i\omega_{l}-\xi_{0\bm{k}})\sigma_{0}$, and $\xi_{0\bm{k}}$=$\xi_{0-\bm{k}}$ is the single-particle dispersion for the double degenerate bands.
The mean-field free energy can be expressed as
\begin{equation}\label{eq:mffe}
\begin{split}
&\mathcal{F}\equiv{}-\frac{1}{\beta}\mbox{Tr\ ln} \mathcal{G}^{-1} = \mbox{const.} - \frac{1}{\beta}\mbox{Tr\ ln}(1+\mathcal{G}_{0}\Sigma) = \mbox{const.} + \sum_{j=1}^{\infty}\mathcal{F}^{(2j)},\\
&\mathcal{F}^{(2j)}=\frac{1}{2j\beta}\sum_{l,\bm{k}}\mbox{Tr}[(\mathcal{G}_{0}(i\omega_{l},\bm{k})\Sigma(\bm{k}))^{2j}]=\frac{1}{j\beta}\sum_{l,\bm{k}}\mbox{Tr}[(-G_{0}(i\omega_{l},\bm{k})\Delta_{\bm{k}}G_{0}(-i\omega_{l},-\bm{k})\Delta_{\bm{k}}^{\dagger})^{j}].
\end{split}
\end{equation}
Please note that in the first line of Eq~\eqref{eq:mffe} we temporarily omit the sum over $\omega_{l}$ and $\bm{k}$, as well as all the indices of $\omega_{l}$ and $\bm{k}$.

Then we focus on the two component pairing functions in the irreducible representation $E$, i.e., $\Delta_{\bm{k}}$=$i\left[\sum_{m=1}^{2}\phi_{m}\left(\psi_{m\bm{k}}+\bm{d}_{m\bm{k}}\cdot\bm{\sigma}\right)\right]\sigma_{y}$.
For $j=1$, we obtain
\begin{equation}
\begin{split}
\mathcal{F}^{(2)} & = -\frac{1}{\beta}\sum_{l,\bm{k}}\mbox{Tr}[G_{0}(i\omega_{l},\bm{k})\Delta_{\bm{k}}G_{0}(-i\omega_{l},-\bm{k})\Delta_{\bm{k}}^{\dagger}]= \frac{1}{\beta}\sum_{l,\bm{k}}\frac{\mbox{Tr}[\Delta_{\bm{k}}\Delta_{\bm{k}}^{\dagger}]}{(i\omega_{l}-\xi_{0\bm{k}})(i\omega_{l}+\xi_{0\bm{k}})}\\
& = \frac{1}{\beta}\sum_{l,\bm{k}}\frac{\mbox{Tr}[\sigma_{0}\left((\psi_{1\bm{k}}^{2}+d_{1\bm{k}}^{2})|\phi_{1}|^{2}+(\psi_{2\bm{k}}^{2}+d_{2\bm{k}}^{2})|\phi_{2}|^{2}+(\psi_{1\bm{k}}\psi_{2\bm{k}}+\bm{d}_{1\bm{k}}\cdot\bm{d}_{2\bm{k}})(\phi_{1}\phi_{2}^{*}+\phi_{1}^{*}\phi_{2})\right)]}{(i\omega_{l}-\xi_{0\bm{k}})(i\omega_{l}+\xi_{0\bm{k}})}.	\\
\end{split}
\end{equation}
The momentum summation of $(\phi_{1}\phi_{2}^{*}+\phi_{1}^{*}\phi_{2})$ terms in the last line will vanish because it is odd under $k_{x}\rightarrow-k_{x}$. Thus only $|\phi_{1}|^{2}$ and $|\phi_2|^{2}$ contribute to the quadratic order. For the quartic order, we have
\begin{equation}
\begin{split}
\mathcal{F}^{(4)} & =
\frac{1}{2\beta}\sum_{l,\bm{k}}\mbox{Tr}\left[\left(G_{0}(i\omega_{l},\bm{k})\Delta_{\bm{k}}G_{0}(-i\omega_{l},-\bm{k})\Delta_{\bm{k}}^{\dagger}\right)^{2}\right]= \frac{1}{2\beta}\sum_{l,\bm{k}}\frac{\mbox{Tr}[\Delta_{\bm{k}}\Delta_{\bm{k}}^{\dagger}\Delta_{\bm{k}}\Delta_{\bm{k}}^{\dagger}]}{(i\omega_{l}-\xi_{0\bm{k}})^{2}(i\omega_{l}+\xi_{0\bm{k}})^{2}}\\
&=\frac{1}{\beta}\sum_{l,\bm{k}}\frac{1}{(\omega_{l}^{2}+\xi_{0\bm{k}}^{2})^{2}}\Big{(}\left((\psi_{1}^{2}+d_{1}^{2})^{2}+4\psi_{1}^{2}d_{1}^{2}\right)|\phi_{1}|^{4}+\left((\psi_{2}^{2}+d_{2}^{2})^{2}+4\psi_{2}^{2}d_{2}^{2}\right)|\phi_{2}|^{4} \\
&+2\left((\psi_{1}^{2}+d_{1}^{2})(\psi_{2}^{2}+d_{2}^{2})+2(\psi_{1}\psi_{2}+\bm{d}_{1}\cdot\bm{d}_{2})^{2}+2|\psi_{1}\bm{d}_{2}+\psi_{2}\bm{d}_{1}|^{2}\right)|\phi_{1}\phi_{2}|^{2}\\ 
&+((\psi_{1}\psi_{2}+\bm{d}_{1}\cdot\bm{d}_{2})^{2}+|\psi_{1}\bm{d}_{2}+\psi_{2}\bm{d}_{1}|^{2}-|\bm{d}_{1}\times{}\bm{d}_{2}|^{2})\left(\phi_{1}\phi_{2}^{*}-\phi_{1}^{*}\phi_{2}\right)^{2}\Big{)},
\end{split}
\end{equation}
where the subindex $\bm{k}$ of the last three lines is omitted. We discard the terms containing odd-order of $\psi_{1\bm{k}}$($\psi_{2\bm{k}}$) and $\bm{d}_{1\bm{k}}$($\bm{d}_{2\bm{k}}$), such as $\psi_{1\bm{k}}\psi_{2\bm{k}}\bm{d}_{1\bm{k}}\cdot\bm{d}_{2\bm{k}}|\phi_{1}\phi_{2}|^{2}$, since they vanish after the momentum summation.

Regarding the $C_{3v}$ symmetry and Eq.~(3) in the main text, we obtain
\begin{equation}
\beta_{2} = \frac{1}{\beta} \sum_{\omega_{l},\bm{k}}\frac{(\psi_{1\bm{k}}\psi_{2\bm{k}}+\bm{d}_{1\bm{k}}\cdot\bm{d}_{2\bm{k}})^{2}+|\psi_{1\bm{k}}\bm{d}_{1\bm{k}}+\psi_{2\bm{k}}\bm{d}_{2\bm{k}}|^{2}-|\bm{d}_{1\bm{k}}\times{}\bm{d}_{2\bm{k}}|^{2}}{(\omega_{l}^{2}+\xi_{0\bm{k}}^{2})^{2}}.
\end{equation}
Apparently, for pure spin-singlet states,
\begin{equation}
\beta_{2} = \frac{1}{\beta} \sum_{\bm{k}}\frac{1-2n_{F}(\xi_{0\bm{k}})}{2\xi_{0\bm{k}}}(\psi_{1\bm{k}}\psi_{2\bm{k}})^{2}\approx{}C\langle(\psi_{1\bm{k}}\psi_{2\bm{k}})^{2}\rangle_{\mathrm{FS}},
\end{equation}
where $n_{F}$ is the Fermi-Dirac distribution function,
$C\equiv(1-2n_{F}(\xi_{0\bm{k}}))/{2\xi_{0\bm{k}}}\approx\beta/4>0$. 
The last step of Eq.~(13) can only be obtained in the low-temperature limit. 
Similarly, for pure spin-triplet states, we have
\begin{equation}
\beta_{2} = \sum_{\bm{k}}\frac{1-2n_{F}(\xi_{0\bm{k}})}{2\xi_{0\bm{k}}} \left[(\bm{d}_{1\bm{k}}\cdot\bm{d}_{2\bm{k}})^{2}-|\bm{d}_{1\bm{k}}\times{}\bm{d}_{2\bm{k}}|^{2}\right]\approx{}C\langle(\bm{d}_{1\bm{k}}\cdot\bm{d}_{2\bm{k}})^{2}-|\bm{d}_{1\bm{k}}\times{}\bm{d}_{2\bm{k}}|^{2}\rangle_{\mathrm{FS}}.
\end{equation}

\section{Proof of the Lemma}
We shall derive the Ginzburg-Landau free energy for a system with strong SOC. Define the energy dispersion measured from the helical Fermi surface, $\xi_{\bm{k}\pm}$=$\epsilon_{\bm{k}\pm}-\mu$, and introduce $\xi_{\bm{k}}$=$(\xi_{\bm{k}+}+\xi_{\bm{k}-})/2$ and $\delta_{\bm{k}}$=$(\xi_{\bm{k}+}-\xi_{\bm{k}-})/2$ ({\bf $\delta_{\bm{k}}\sim{}\Delta_{SOC}$}), we can write down the pairing function and Matsubara Green's function in the pseudospin  basis $|\bm{k},\pm\rangle$ respectively as follows, 
\begin{equation}
\Delta(\bm{k})=\left(\begin{array}{ll}
\Delta_{++}(\bm{k}) &\Delta_{+-}(\bm{k}) \\
\Delta_{-+}(\bm{k}) &\Delta_{--}(\bm{k})
\end{array}\right),
\end{equation}
and
\begin{equation}
\begin{split}
&\tilde{G}^{-1}(i\omega_{l},\bm{k})=i\omega_{l}-\left(\begin{array}{ll}
\xi_{\bm{k}+} &0 \\
0 &\xi_{\bm{k}-}
\end{array}\right)\equiv\left(\begin{array}{ll}
a_{\bm{k}l}^{-1} &0 \\
0 &b_{\bm{k}l}^{-1}
\end{array}\right),\\
\end{split}
\end{equation}
where $a_{\bm{k}l}^{-1}=i\omega_{l}-\xi_{\bm{k}+}$ and $b_{\bm{k}l}^{-1}=i\omega_{l}-\xi_{\bm{k}-}$. Here we consider the free energy with quadratic-order and quartic-order. According to Eq.~\eqref{eq:mffe}, the quadratic-order free energy can be expressed as
\begin{equation}
\begin{split}
\mathcal{F}^{(2)}&=-\frac{1}{\beta}\sum_{l,\bm{k}}\mbox{Tr}[\tilde{G}(i\omega_{l},\bm{k})\Delta({\bm{k}})\tilde{G}(-i\omega_{l},-\bm{k})\Delta({\bm{k}})]\\
&=-\frac{1}{\beta}\sum_{n,\bm{k}}\left(|a_{\bm{k}l}|^{2}|\Delta_{++}(\bm{k})|^{2}+(a_{\bm{k}l}b_{\bm{k}l}^{*}+a_{\bm{k}l}^{*}b_{\bm{k}l})|\Delta_{+-}(\bm{k})|^{2}+|b_{\bm{k}l}|^{2}|\Delta_{--}(\bm{k})|^{2} \right),
\end{split}
\end{equation}
where $\Delta_{+-}(\bm{k})=-\Delta_{-+}(-\bm{k})$, $a_{\bm{k}l}$=$a_{-\bm{k}l}$ and $b_{\bm{k}l}$=$b_{-\bm{k}l}$ have been used. By summing over Matsubara frequency $i\omega_{l}$, we obtain
\begin{equation}
\begin{split}
&\sum_{l}|a_{\bm{k}l}|^{2}=\frac{\beta}{2\xi_{\bm{k}+}}(n_{F}(-\xi_{\bm{k}+})-n_{F}(\xi_{\bm{k}+})),\\
&\sum_{l}|b_{\bm{k}l}|^{2}=\frac{\beta}{2\xi_{\bm{k}-}}(n_{F}(-\xi_{\bm{k}-})-n_{F}(\xi_{\bm{k}-})),\\
&\sum_{l}(a_{\bm{k}l}b_{\bm{k}l}^{*}+a^{*}_{\bm{k}l}b_{\bm{k}l})=\frac{\beta}{2\xi_{\bm{k}}}(1-n_{F}(\xi_{\bm{k}+})-n_{F}(\xi_{\bm{k}-})).
\end{split}
\end{equation}
It is easy to show that when $\bm{k}$ is close to the Fermi surface,
\begin{equation}
\begin{split}
&\lim\limits_{\xi_{\bm{k}+}=0}\left(\sum_{l}|a_{\bm{k}l}|^{2}\right)=\lim\limits_{\xi_{\bm{k}-}=0}\left(\sum_{l}|b_{\bm{k}l}|^{2}\right)=\frac{\beta^{2}}{4},\\
&\lim\limits_{\xi_{\bm{k}+}=0}\left(\sum_{l}a_{\bm{k}l}^{*}b_{\bm{k}l}\right)\sim{}\frac{\beta}{4\delta_{\bm{k}}}.\\
\end{split}
\end{equation}
In the region of $\delta_{\bm{k}}\gg\Delta>{}k_{B}T$, we have
\begin{equation}
\frac{\sum_{l}(a_{\bm{k}l}b_{\bm{k}l}^{*}+a^{*}_{\bm{k}l}b_{\bm{k}l})}{\sum_{l}|a_{\bm{k}l}|^{2}}\sim{}\frac{1}{\beta\delta_{\bm{k}}} \ll 1,
\end{equation}
which indicates that the term containing $\Delta_{+-}(\bm{k})$ is not important. Finally we have
\begin{equation}
\begin{split}
\mathcal{F}^{(2)}\approx-\frac{1}{\beta}\sum_{l,\bm{k}}\left(|a_{\bm{k}l}|^{2}|\Delta_{++}(\bm{k})|^{2}+|b_{\bm{k}l}|^{2}|\Delta_{--}(\bm{k})|^{2} \right).
\end{split}
\end{equation}

The free energy from the quartic-order is
\begin{equation}
\begin{split}
\mathcal{F}^{(4)} =&\frac{1}{2\beta}\sum_{l,\bm{k}}\left\{|a_{l}|^{4}|\Delta_{++}|^{4}+|b_{l}|^{4}|\Delta_{--}|^{4}\right\}\\ 
&+\frac{1}{2\beta}\sum_{l,\bm{k}}\left\{2(|a_{l}|^{2}|\Delta_{++}|^{2}+|b_{l}|^{2}|\Delta_{--}|^{2})(a_{l}b_{l}^{*}|\Delta_{+-}|^{2}+a_{l}^{*}b_{l}|\Delta_{-+}|^{2})\right\}\\
&+\frac{1}{2\beta}\sum_{l,\bm{k}}\left\{\left[(a_{l}b_{l}^{*})^{2}+(a_{l}^{*}b_{l})^{2}\right]|\Delta_{+-}|^{4}+2|a_{l}b_{l}|^{2}\left(\Delta_{++}\Delta_{--}\Delta_{+-}^{*}\Delta_{-+}^{*}+c.c\right)\right\},\\
\end{split}
\end{equation}
where the indices $\bm{k}$ for $a_{l}(b_{l})$ and $\Delta_{\pm\pm}$ are omitted. Again, by summing over $i\omega_{l}$,  we obtain
\begin{equation}
\begin{split}
&\sum_{l}|a_{\bm{k}l}|^{4}=\frac{\beta}{4\xi_{\bm{k}+}^{2}}\left\{ -2\beta{}n_{F}(\xi_{\bm{k}+})n_{F}(-\xi_{\bm{k}+}) + \frac{1}{\xi_{\bm{k}+}}(n_{F}(-\xi_{\bm{k}+})-n_{F}(\xi_{\bm{k}+}))\right\},\\
&\sum_{l}|b_{\bm{k}l}|^{4}=\frac{\beta}{4\xi_{\bm{k}-}^{2}}\left\{ -2\beta{}n_{F}(\xi_{\bm{k}-})n_{F}(-\xi_{\bm{k}-}) + \frac{1}{\xi_{\bm{k}-}}(n_{F}(-\xi_{\bm{k}-})-n_{F}(\xi_{\bm{k}-}))\right\},\\
&\sum_{l}|a_{\bm{k}l}b_{\bm{k}l}|^{2}=\frac{\beta}{4\xi_{\bm{k}-}\xi_{\bm{k}+}}\left\{\frac{n_{F}(-\xi_{\bm{k}-})-n_{F}(\xi_{\bm{k}+})}{\xi_{\bm{k}}}+\frac{n_{F}(\xi_{\bm{k}+})-n_{F}(\xi_{\bm{k}-})}{\delta_{\bm{k}}}\right\},\\
&\sum_{l}|a_{\bm{k}l}|^{2}a_{\bm{k}l}^{*}b_{\bm{k}l}=\frac{\beta}{4\xi_{\bm{k}}\xi_{\bm{k}+}}\left\{\frac{n_{F}(-\xi_{\bm{k}+})-n_{F}(\xi_{\bm{k}-})}{2\xi_{\bm{k}}}+\frac{n_{F}(\xi_{\bm{k}+})-n_{F}(\xi_{\bm{k}-})}{2\delta_{\bm{k}}}\right\}\\
&\qquad\qquad\qquad\qquad\quad+\frac{\beta}{4\xi_{\bm{k}}\xi_{\bm{k}+}}\left\{ -\beta{}n_{F}(\xi_{\bm{k}+})n_{F}(-\xi_{\bm{k}+}) + \frac{n_{F}(-\xi_{\bm{k}+})-n_{F}(\xi_{\bm{k}+})}{2\xi_{\bm{k}+}}\right\},\\
&\sum_{l}(a_{\bm{k}l}^{*}b_{\bm{k}l})^{2}=\frac{\beta}{4\xi_{\bm{k}}^{2}}\left\{-\beta(n_{F}(\xi_{\bm{k}+})n_{F}(-\xi_{\bm{k}+})+n_{F}(\xi_{\bm{k}-})n_{F}(-\xi_{\bm{k}-}))+\frac{n_{F}(-\xi_{\bm{k}+})-n_{F}(\xi_{\bm{k}-})}{\xi_{\bm{k}}}\right\}.\\
\end{split}
\end{equation}
The following equations are valid that
\begin{equation}
\begin{split}
&\lim\limits_{\xi_{\bm{k}+}=0}\left(\sum_{l}|a_{\bm{k}l}|^{4}\right)=\lim\limits_{\xi_{\bm{k}-}=0}\left(\sum_{l}|b_{\bm{k}l}|^{4}\right)=\frac{\beta^{4}}{12},\\
&\lim\limits_{\xi_{\bm{k}+}=0}\left(\sum_{l}|a_{\bm{k}l}b_{\bm{k}l}|^{2}\right)=\lim\limits_{\xi_{\bm{k}-}=0}\left(\sum_{l}|a_{\bm{k}l}b_{\bm{k}l}|^{2}\right)=\frac{\beta^{2}}{16\delta_{\bm{k}}^{2}}+O\left(\frac{\beta}{\delta_{\bm{k}}^{3}}\right),\\
&\lim\limits_{\xi_{\bm{k}+}=0}\left(\sum_{l}|a_{\bm{k}l}|^{2}a_{\bm{k}l}^{*}b_{\bm{k}l}\right)=\frac{\beta^{2}}{16\delta_{\bm{k}}^{2}}+O\left(\frac{\beta}{\delta_{\bm{k}}^{3}}\right),\\
&\lim\limits_{\xi_{\bm{k}+}=0}\left(\sum_{l}(a_{\bm{k}l}^{*}b_{\bm{k}l})^{2}\right)\sim{}\frac{\beta^{2}}{\delta_{\bm{k}}^{2}}.\\
\end{split}
\end{equation}
So, the total free energy consisting of the quadratic and quartic order is
\begin{equation}
\mathcal{F}\approx\mbox{const}.-\frac{1}{\beta}\sum_{l,\bm{k}}\left(|a_{\bm{k}l}\Delta_{++}(\bm{k})|^{2}+|b_{\bm{k}l}\Delta_{--}(\bm{k})|^{2} - \frac{|a_{\bm{k}l}\Delta_{++}(\bm{k})|^{4}+|b_{\bm{k}l}\Delta_{--}(\bm{k})|^{4}}{2}\right).
\end{equation}

\section{More information about quasiparticle gap structures}

For the irreducible representation $E$ of $C_{3v}$, the corresponding gap function is written as $\Delta(\bm{k})$=$i\left[\sum_{m=1}^{2}\phi_{m}\left(\epsilon_{1}\psi^{E}_{m,\bm{k}}+(\epsilon_{2}\bm{d}^{E,xy}_{m,\bm{k}}+\epsilon_{3}\bm{d}^{E,z}_{m,\bm{k}})\cdot\bm{\sigma}\right)\right]\sigma_{y}$. 
In the pseudospin  basis, the inter-band pairings are not important, thus we only consider the intra-band pairings $\Delta_{ss}(\bm{k})$ with $s=\pm$.

In the case of $(\phi_{1},\phi_{2})=\phi(1,1)$, the modulus of gap functions reads
\begin{equation}
|\Delta_{++}(\bm{k})|=k|(\epsilon_{1}k-\epsilon_{2}u_{\bm{k}})(\sin(2\theta_{\bm{k}})+\cos(2\theta_{\bm{k}}))-2\epsilon_{3}\tilde{v}_{\bm{k}}\cos(3\theta_{\bm{k}})(\sin\theta_{\bm{k}}+\cos\theta_{\bm{k}})|,
\end{equation}
where $v_{\bm{k}}=2\tilde{v}_{\bm{k}}\cos(3\theta_{\bm{k}})$, and the expression of  $u_{\bm{k}}$ and $v_{\bm{k}}$ can be found in the Table.~II in the main text.
It is easy to see that when $\epsilon_{3}=0$, $|\Delta_{++}(\bm{k})|=0$ for $\sin(2\theta_{\bm{k}})=\pm{}1/\sqrt{2}$. 
When $\epsilon_{3}\neq0$,  $|\Delta_{++}(\bm{k})|=0$ requires that
\begin{equation}
\begin{split}
&\frac{(\epsilon_{1}k-\epsilon_{2}u_{\bm{k}})}{2\epsilon_{3}\tilde{v}_{\bm{k}}}=Y(\theta_{\bm{k}}),\\ 
\end{split}
\end{equation}
where $Y(\theta_{\bm{k}})=\cos(3\theta_{\bm{k}})(\sin\theta_{\bm{k}}+\cos\theta_{\bm{k}})/(\sin(2\theta_{\bm{k}})+\cos(2\theta_{\bm{k}}))$.
Because $u_{\bm{k}}$ and $\tilde{v}_{\bm{k}}$ vary much more slowly than $Y(\theta_{\bm{k}})$ on the Fermi surface, the left side of above formula can be treated as a constant. 
Note that	$\sin(\theta_{\bm{k}})=\pm\ sqrt{2}/2\Longrightarrow{}Y(\theta_{\bm{k}})=0$ and $\sin(2\theta_{\bm{k}})=\pm\sqrt{2}/2\Longrightarrow{}Y(\theta_{\bm{k}})\rightarrow\infty$, so the domain of $Y(\theta_{\bm{k}})$ is $[-\infty,\infty]$, which indicates that the condition of $|\Delta_{++}(\bm{k})|=0$ can always be satisfied and nodes are stable for the TRS phase.

In the case of $(\phi_{1},\phi_{2})=\phi(1,i)$, it is easy to verify that 
\begin{equation}\label{eq:gspip}
\begin{split}
|\Delta_{++}(\bm{k})|&=|k\left((\epsilon_{1}k-\epsilon_{2}u_{\bm{k}})e^{i\theta_{\bm{k}}}-2\epsilon_{3}\tilde{v}_{\bm{k}}\cos(3\theta_{\bm{k}}) \right)|\\
&=k\sqrt{(\epsilon_{1}k-\epsilon_{2}u_{\bm{k}})^{2}-4\epsilon_{3}\tilde{v_{\bm{k}}}(\epsilon_{1}k-\epsilon_{2}u_{\bm{k}})\cos\theta_{\bm{k}}\cos(3\theta_{\bm{k}})+4\epsilon_{3}^{2}\tilde{v}^{2}_{\bm{k}}\cos^{3}(3\theta_{\bm{k}})}.
\end{split}
\end{equation}
It is obvious the modulus of gap function never vanishes for the TRS breaking phase unless $\epsilon_{1}=\epsilon_{2}=0$. And $\cos(3\theta_{\bm{k}})\cos(\theta_{\bm{k}})$ breaks the six-fold symmetry but preserves the two-fold symmetry.
Note that the physics of $\Delta_{--}(\bm{k})$ is the same as $\Delta_{++}(\bm{k})$.

\end{widetext}

\end{document}